\documentclass[10pt, conference]{IEEEtran}
\IEEEoverridecommandlockouts
\usepackage{cite}
\usepackage{amsmath,amssymb,amsfonts}
\usepackage{graphicx}
\usepackage{textcomp}
\usepackage{xcolor}
\usepackage{algorithm}
\usepackage{caption}
\usepackage{upgreek}
\usepackage{multirow}
\usepackage{algpseudocode}
\usepackage{tablefootnote}
\usepackage{bm}
\usepackage{tabularray}
\usepackage{threeparttable}
\usepackage{makecell}
\usepackage{tabularx} 
\usepackage{booktabs} 
\usepackage{array} 
\usepackage{ragged2e}
\usepackage{graphicx}
\usepackage{subcaption} 
\usepackage{amsmath}

\newcolumntype{M}[1]{>{\centering\arraybackslash}m{#1}}
\newcolumntype{C}[1]{>{\centering\arraybackslash}p{#1}}
\def\BibTeX{{\rm B\kern-.05em{\sc i\kern-.025em b}\kern-.08em
    T\kern-.1667em\lower.7ex\hbox{E}\kern-.125emX}}
\begin{document}

\title{APSQ: Additive Partial Sum Quantization with Algorithm-Hardware Co-Design\\
\thanks{*Both authors contributed equally to this research.}
\thanks{$^\dag$Corresponding author.}
}
\author{Yonghao Tan*, Pingcheng Dong*, Yongkun Wu, Yu Liu, Xuejiao Liu, Peng Luo, \\
Shih-Yang Liu, Xijie Huang, Dong Zhang, Luhong Liang, Kwang-Ting Cheng$^\dag$ \\
The Hong Kong University of Science and Technology, AI Chip Center for Emerging Smart Systems (ACCESS)}


\maketitle

\begin{abstract}
DNN accelerators, significantly advanced by model compression and specialized dataflow techniques, have marked considerable progress. However, the frequent access of high-precision partial sums (PSUMs) leads to excessive memory demands in architectures utilizing input/weight stationary dataflows. Traditional compression strategies have typically overlooked PSUM quantization, which may account for 69\% of power consumption. This study introduces a novel Additive Partial Sum Quantization (APSQ) method, seamlessly integrating PSUM accumulation into the quantization framework. A grouping strategy that combines APSQ with PSUM quantization enhanced by a reconfigurable architecture is further proposed. The APSQ performs nearly lossless on NLP and CV tasks across BERT, Segformer, and EfficientViT models while compressing PSUMs to INT8. This leads to a notable reduction in energy costs by 28-87\%. Extended experiments on LLaMA2-7B demonstrate the potential of APSQ for large language models. Code is available at \emph{https://github.com/Yonghao-Tan/APSQ}.
\end{abstract}

\begin{IEEEkeywords}
Hardware-aware quantization, partial sum quantization, dataflow, DNN, Transformer
\end{IEEEkeywords}

\section{Introduction}
Recent advancements in deep neural networks (DNNs) have driven substantial progress in computer vision (CV) and natural language processing (NLP) tasks, including image classification \cite{dosovitskiy2020image}, semantic segmentation \cite{Cai_2023_ICCV, xie2021segformer}, object detection \cite{quan2023centralized}, and question answering \cite{devlin2018bert}. Furthermore, the rise of generative large language models (LLMs) \cite{touvron2023llama} has profoundly influenced various applications. Central to this advancement is the Transformer architecture \cite{vaswani2017attention}, renowned for its self-attention mechanism that effectively captures long-range dependencies. However, Transformers are notorious for their intensive computation and memory requirements. For instance, semantic segmentation for self-driving vehicles involves processing over 20,000 tokens, while BERT models must handle tokens with substantial hidden dimensions, such as 4,096, resulting in considerable memory overhead. These challenges significantly impact hardware deployment and acceleration. To address these obstacles, the algorithm community has developed model compression techniques, including quantization \cite{jacob2018quantization, kim2021bert, DBLP:conf/iclr/EsserMBAM20, liu2023oscillation}, pruning \cite{liu2019metapruning}, and distillation \cite{shu2021channel}.

To further accelerate DNNs at the hardware level, prior works have focused on designing dedicated accelerators utilizing diverse dataflows \cite{chen2014diannao, chen2016eyeriss, tu2017deep, you2023vitcod, lu2021sanger}. They can be categorized into Input Stationary (IS), Weight Stationary (WS), and Output Stationary (OS). The IS reduces data movement by keeping input tiles stationary \cite{chen2016eyeriss} to achieve minimum access to the input buffer. The WS tries to minimize weight movement from external DRAM due to insufficient on-chip weight buffer \cite{tu2017deep}. The OS achieves optimization by reusing partial sum (PSUM) in the output registers \cite{chen2014diannao}, enhancing computational efficiency in scenarios with extensive accesses of PSUMs.  In addition, Chen et al. \cite{chen2016eyeriss} proposed Row Stationary (RS) dataflow, which aims to minimize the energy consumption of data movement for convolutional neural networks (CNNs). However, since Transformers rarely employ standard convolution, the RS dataflow has not been widely adopted. The effectiveness of these dataflows is contingent upon several critical aspects, including the layer configuration, degree of parallelism, and on-chip SRAM size. Addressing this, Tu et al. \cite{tu2017deep} proposes a reconfigurable architecture specifically designed to leverage the unique characteristics of these diverse dataflows optimally. Besides, recent Transformer accelerators \cite{you2023vitcod, lu2021sanger} employ Score/Key stationary dataflows to accelerate the attention, which is aligned with the dataflows introduced above.

\begin{figure}[!t]
  \centering
    \captionsetup{skip=1mm}
    \includegraphics[width=1.0\linewidth]{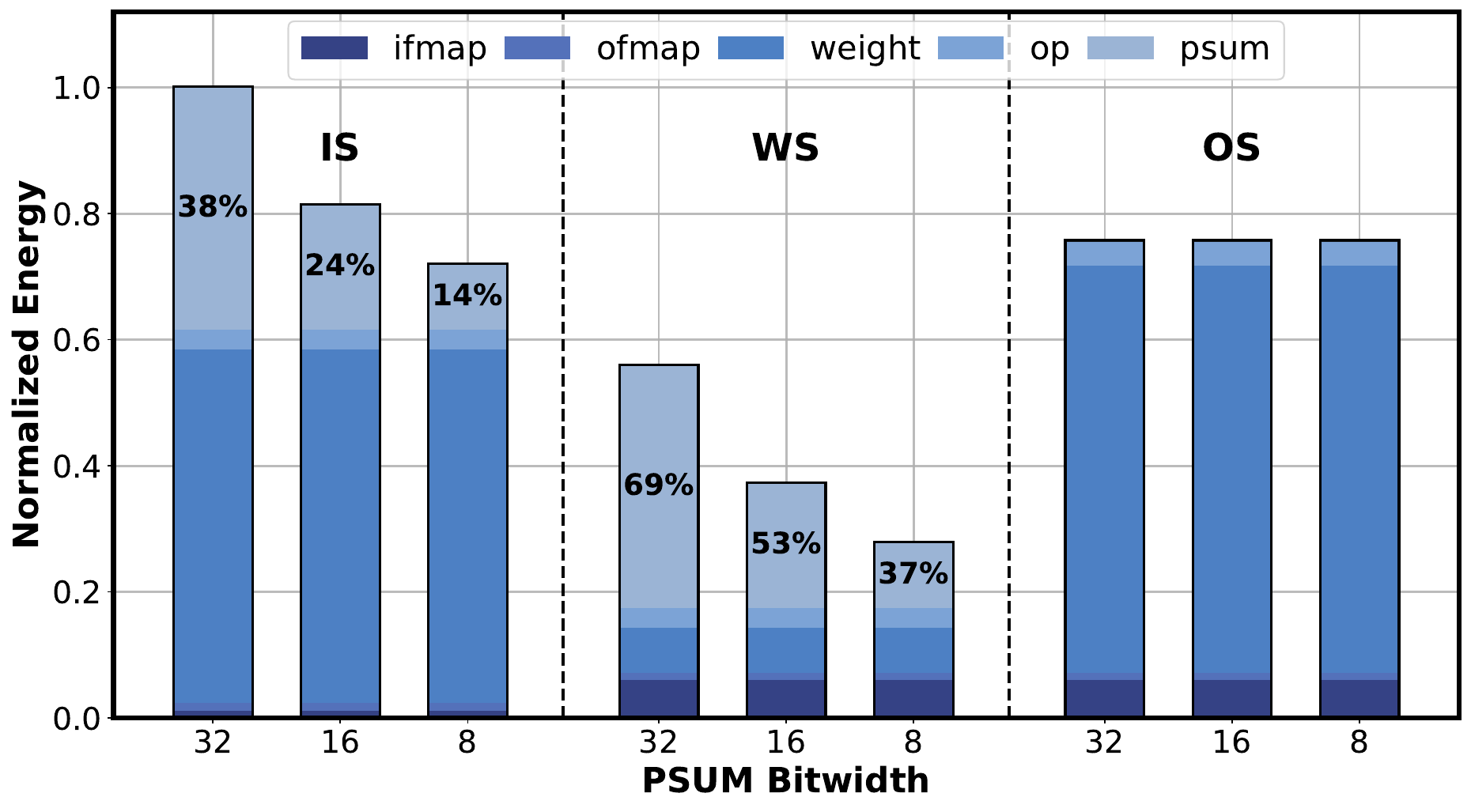}
    \caption{Energy breakdown of IS, WS, and OS dataflows for the BERT-Base model with 128 input token-length.}
    \label{Figure1}
    \vspace{-4mm}  
\end{figure}

The key distinction among the aforementioned dataflows lies in the fact that the IS and WS frequently store and fetch PSUMs from SRAM/DRAM for accumulation, while OS updates PSUMs directly within low-cost registers. We perform an energy consumption analysis following the methods adopted in \cite{tu2017deep} as shown in Fig. \ref{Figure1}. It can be seen that PSUMs constitute a large portion of the energy for IS and WS compared with input feature map (ifmap), output feature map (ofmap), weight, and multiply-add operations (op), especially with larger PSUM bit-width, up to 69\% of total power consumption. Since PSUMs are obtained by matrix accumulation, they need to be stored in higher precision (e.g., INT16/32) than the quantized weights/activation that are stored in INT8 or lower. Recent studies \cite{azamat2023partial, bai2023partial} have proposed PSUM quantization (PSQ) technique to reduce analog-to-digital converter (ADC) overheads of high-precision PSUMs in ReRAM-based accelerator. Nevertheless, the PSUMs after ADC are dequantized to their original bit-width and stored in on-chip SRAM, where the memory access overhead of PSUMs fails to be alleviated.

\begin{figure*}[h]
  \centering
     \captionsetup{skip=1mm}
    \includegraphics[width=0.95\linewidth]{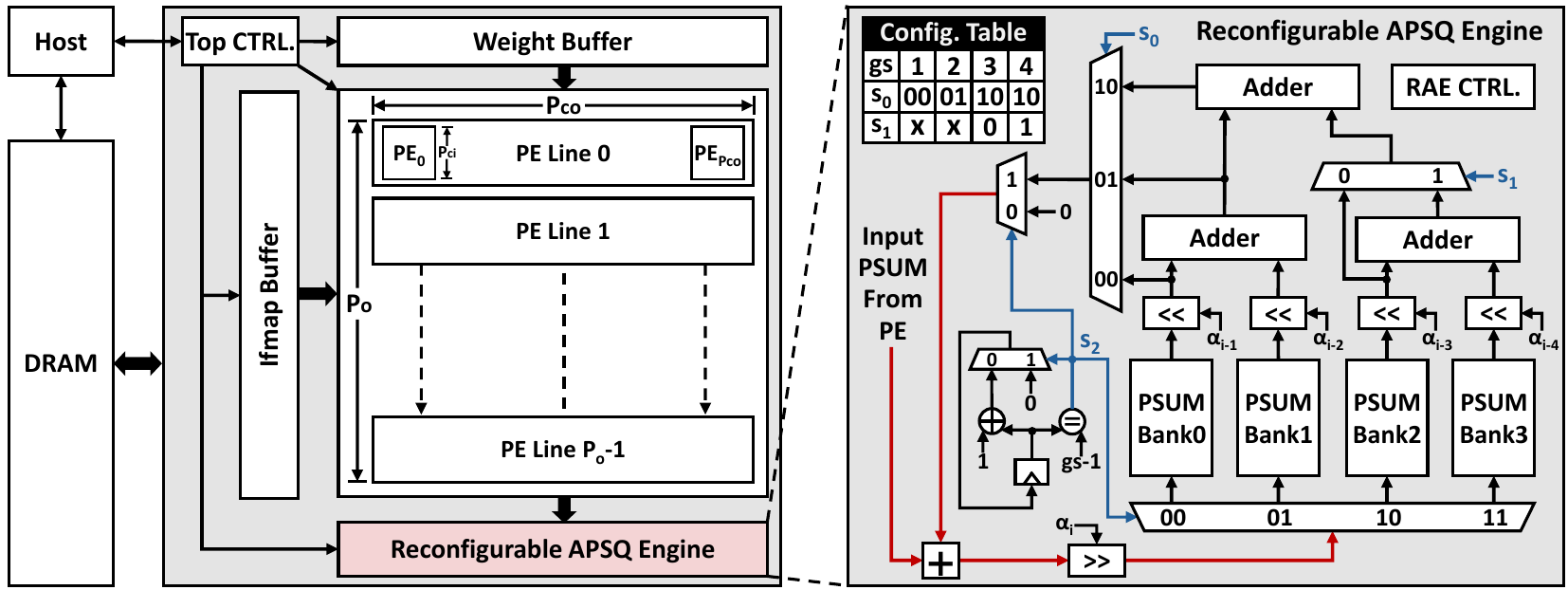}
    \caption{The analytical DNN accelerator and the Reconfigurable APSQ Engine (RAE) in this work.}
    \vspace{-4mm}  
    \label{Figure2}
\end{figure*}

In this study, we introduce a novel additive PSUM quantization strategy denoted as APSQ, which, to the best of our knowledge, is the first to address challenges in IS and WS encountered with large PSUMs. The APSQ enables each quantizer to consider the cumulative prior outcomes rather than just focusing on the current PSUM. In addition, we propose a grouping strategy to improve accuracy with a reconfigurable architecture to support various grouping configurations. The main contributions of this paper are summarized as follows:
\begin{itemize}
\item We reveal the critical role of high-precision PSUM access and storage in both IS and WS dataflows, refining an analytical framework with PSUM precision awareness.
\item A novel method termed APSQ is designed by introducing the accumulation effect into the quantizer, thereby achieving INT8 PSUM storage and accesses.
\item We propose a grouping strategy integrating PSUM quantization and APSQ within a reconfigurable architecture, efficiently accommodating various group sizes with minimal additional hardware resources.
\item The INT8 APSQ minimizes accuracy loss to 0.16\% for BERT-Base, 0.61\% and 0.83\% for Segformer-B0 and EfficientViT-B1, while the energy costs are saved by 28-87\% for the IS and WS. We further apply APSQ for LLaMA2-7B and achieved only 0.59\% accuracy drop while providing up to \( 31.7\times \) energy savings, highlighting its significant potential for LLMs.

\end{itemize}

\section{Preliminary}
\subsection{Energy Efficiency Analysis}
\label{2.1}
In the realm of DNN accelerators, a key factor is energy efficiency in terms of operations/watt, significantly influenced by precision, layer size, dataflow, and architecture. This study adopts the methodologies from \cite{chen2016eyeriss, tu2017deep} to examine energy efficiency in a typical DNN accelerator system shown in Fig. 2. This system includes a multiply-accumulate (MAC) array, on-chip SRAM, off-chip DRAM, and a top controller. The MAC array is organized based on $P_{o}$, $P_{ci}$, and $P_{co}$, which define parallelism in the output feature map, input channels, and output channels, respectively. The on-chip SRAM comprises a 128KB input/output buffer and a 64KB weight buffer, while the off-chip DRAM capacity is assumed sufficient, and the top controller manages module configuration. Following \cite{tu2017deep}, the total energy cost $E_{total}$ is:
\begin{equation}
    E_{total}=N_{d}\cdot E_{dram}+N_{s}\cdot E_{sram}+N_{m}\cdot E_{mac}
\end{equation}
where the $E_{dram}$, $E_{sram}$, and $E_{mac}$ represent the energy cost for each access to DRAM and SRAM, as well as for a single  MAC operation, respectively. The $N_{d}$, $N_{s}$, and $N_{m}$ correspond to the total number of accesses to DRAM and SRAM, and the overall count of MAC operations utilized, with energy cost values taken from \cite{horowitz20141}. Since the number of MACs for a specific DNN is constant, the energy efficiency largely depends on $E_{dram/sram}$ as proved by \cite{tu2017deep}.


In the analytical framework proposed by \cite{chen2016eyeriss, tu2017deep}, weights, activations, and partial sums (PSUMs) are all stored in INT16 format. However, in certain integer-only quantized DNN accelerators, PSUMs may require higher precision, such as INT32, to maintain accuracy \cite{9566303, 9689050}. Consider a neural network where both weights and activations are quantized to INT8 (W8A8). During a multiply-accumulate (MAC) operation, the product of the quantized weight and activation results in an INT16 value. When these products are accumulated along the input channel dimension, denoted as \( C_i \), the accumulation depth plays a crucial role in determining the required bit width to prevent overflow. Specifically, to ensure numerical stability, the PSUM should be stored using \( 16 + \log_2(C_i) \) bits. For instance, in the Feed-Forward Network (FFN) of BERT-Large \cite{devlin2018bert}, where \( C_i = 4096 \) in the MLP layer, this results in a required PSUM precision of up to 28 bits to accommodate the full accumulation range without overflow. Given that memory hierarchy designs are typically byte-based, PSUM is generally stored with 32-bit precision.

To accommodate this variation, we revise the framework by introducing a precision factor as below:
\begin{equation}
    N_{d/s}=S{i}\cdot N_{d/s}^{i}+S_{w}\cdot N_{d/s}^{w}+\beta\cdot S_{o}\cdot N_{d/s}^{p}+S_{o}\cdot N_{d/s}^{o}
\end{equation}
the total number of access to DRAM/SRAM for ifmap, weight, PSUM, and ofmap is defined as $N_{d/s}^{i}$, $N_{d/s}^{w}$, $N_{d/s}^{p}$ and $N_{d/s}^{o}$, with $S_{i}$, $S_{w}$ and $S_{o}$ indicating the size of ifmap, weights, and ofmap. The $\beta$ is the ratio of PSUM precision to that of feature map/weight. For instance, $\beta$ would be 4 if PSUM is in INT32 for an INT8 DNN. In this work, we merely focus on INT8 DNN with WS or IS, since the OS does not struggle with the precision of PSUM.

\subsubsection{Input Stationary} 
The IS-based accelerator keeps the input tiles stationary within the MAC array's registers, facilitating their reuse by weights to update the PSUMs in the output buffer. Similarly, the $N_{s}^{i/w/p/o}$ is summarized in equation (\ref{eq5}):
\begin{equation}
\label{eq5}
\begin{aligned}
&\enspace N_s^w=\left\{
\begin{array}{ll}
1+\left\lceil\frac{H_i}{P_{ih}}\right\rceil\left\lceil\frac{W_i}{P_{iw}}\right\rceil, & S_w<B_w \\
2\left\lceil\frac{H_i}{P_{ih}}\right\rceil\left\lceil\frac{W_i}{P_{iw}}\right\rceil, & S_w \geq B_w
\end{array}\right., 
N_s^i=2 \\
& N_s^p=\left\{\begin{array}{ll}
2\left(\left\lceil\frac{C_{i}}{P_{ci}}\right\rceil-1\right), & \frac{C_o}{P_{co}}\cdot \Tilde{S}_p<B_o \\[1.5mm]
4\left(\left\lceil\frac{C_{i}}{P_{ci}}\right\rceil-1\right), & \frac{C_o}{P_{co}}\cdot \Tilde{S}_p \geq B_o
\end{array}\right.,  N_s^o=2
\end{aligned}
\end{equation}
the weight buffer size is denoted by $B_w$, and the height and width of the enlarged ifmap are $H_i$ and $W_i$, respectively. Besides, the ifmap tile is shaped by $P_{ih}$ and $P_{iw}$ following $P_i = P_{ih} \times P_{iw}$. Notably, the PSUM of IS is smaller than that of WS since the ifmap tile reused by weights is generally small. Consequently, we scale the $\Tilde{S}_p$ by $C_o/P_{co}$. The $N_{p}^{i/w/p/o}$ for DRAM is derived as:
\begin{equation}
\begin{aligned}
&\enspace\enspace N_d^w=\left\{
\begin{array}{ll}
1, & S_w<B_w \\
\left\lceil\frac{H_i}{P_{ih}}\right\rceil\left\lceil\frac{W_i}{P_{iw}}\right\rceil, & S_w \geq B_w
\end{array}\right., 
N_d^i=1 \\
&  N_d^p=\left\{\begin{array}{ll}
0, & \frac{C_o}{P_{co}}\cdot \Tilde{S}_p<B_o \\
2\left(\left\lceil\frac{C_{i}}{P_{ci}}\right\rceil-1\right), & \frac{C_o}{P_{co}}\cdot \Tilde{S}_p \geq B_o
\end{array}\right.,  N_s^o=1
\end{aligned}
\end{equation}

\subsubsection{Weight Stationary} In contrast, the WS-based accelerator loads ifmap tiles to update the PSUMs via reusing $P_{ci}\times P_{co}$ kernel weights, and the ofmap tile is generated after finishing PSUM accumulation. According to this, the components of $N_s$ can be derived as:
\begin{equation}
\begin{aligned}
&\ \, \quad\quad N_s^i=\left\{
\begin{array}{ll}
1+\left\lceil\frac{C_o}{P_{co}}\right\rceil, & \Tilde{S}_i<B_i \\
2\left\lceil\frac{C_o}{P_{co}}\right\rceil, & \Tilde{S}_i \geq B_i
\end{array}\right., 
N_s^w=2 \\
& \enspace N_s^p=\left\{\begin{array}{ll}
2\left(\left\lceil\frac{C_{i}}{P_{ci}}\right\rceil-1\right), & \frac{H_o\cdot W_o}{P_{o}}\cdot\Tilde{S}_p<B_o \\[1.5mm]
4\left(\left\lceil\frac{C_{i}}{P_{ci}}\right\rceil-1\right), & \frac{H_o\cdot W_o}{P_{o}}\cdot\Tilde{S}_p \geq B_o
\end{array}\right.,  N_s^o=2
\end{aligned}
\end{equation}
where $C_i$ and $C_o$ represent the dimensions of input and output channels, the $B_i$ and $B_o$ indicate the capacities of ifmap and ofmap in bytes. The ofmap's height and width are denoted as $H_o$ and $W_o$. Besides, the input tile size $\Tilde{S}i$ is enlarged based on output tiles, kernels, and strides, as explained in \cite{tu2017deep}. The $\Tilde{S}_p$ is the size of tiled PSUM calculated as $\beta\times P_o \times P_{co}$, where $P_{oh}$ and $P_{ow}$ are the height and width of the output/PSUM tile, constrained by $P_o=P_{oh}\times P_{ow}$. If exceeding the $B_i$ or $B_o$, extra read to both the SRAM and DRAM occur, the $N_d^{i/w/p/o}$ is defined as:
\begin{equation}
\begin{aligned}
&\quad\quad\quad\  N_d^i=\left\{
\begin{array}{ll}
1, & \Tilde{S}_i<B_i \\
\left\lceil\frac{C_o}{P_{co}}\right\rceil, & \Tilde{S}_i \geq B_i
\end{array}\right., 
N_d^w=1 \\
&  N_d^p=\left\{\begin{array}{ll}
0, & \frac{H_o\cdot W_o}{P_{o}}\cdot\Tilde{S}_p<B_o \\
2\left(\left\lceil\frac{C_{i}}{P_{ci}}\right\rceil-1\right), & \frac{H_o\cdot W_o}{P_{o}}\cdot\Tilde{S}_p \geq B_o
\end{array}\right.,  N_d^o=1
\end{aligned}
\end{equation}

\subsection{Quantization}
\label{section2.2}
Quantization has emerged as an effective strategy to alleviate computational and memory overheads, particularly advantageous for deployments on hardware accelerators, and can be expressed as:
\begin{equation}
\Tilde{x}=\alpha \cdot Q_k(x)=\alpha \cdot\left\lfloor\operatorname{Clip}\left(\frac{x}{\alpha}, Q_n, Q_p\right)\right\rceil
\end{equation}
\begin{figure}[t] 
  \centering
    \setlength{\abovecaptionskip}{0.cm}
  \begin{subfigure}{\columnwidth}
    \centering
     \captionsetup{skip=1mm}
    \includegraphics[width=\linewidth]{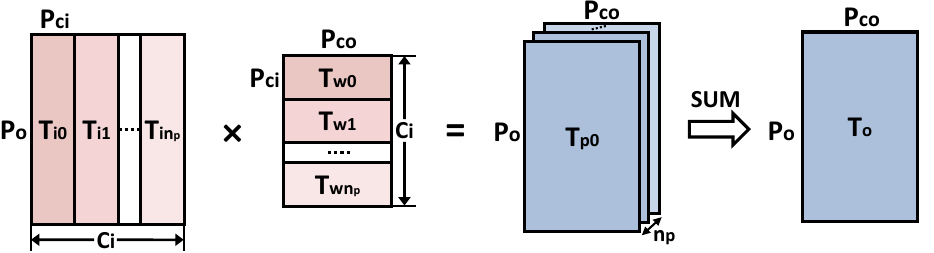}
    \caption{}
    \label{figure3a}
  \end{subfigure}
  \begin{subfigure}{0.46\columnwidth}
    \centering
     \captionsetup{skip=1mm}
    \includegraphics[width=\linewidth]{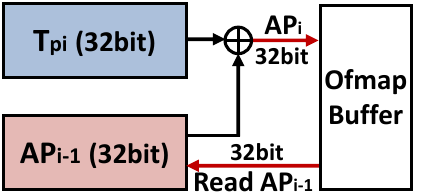}
    \caption{}
    \label{figure3b}
  \end{subfigure}%
  \begin{subfigure}{0.52\columnwidth}
    \centering
     \captionsetup{skip=1mm}
    \includegraphics[width=\linewidth]{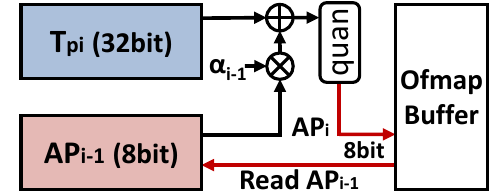}
    \caption{}
    \label{figure3c}
  \end{subfigure}
  \caption{(a) An example of pointwise convolution with tile-based computation scheme, (b) typical partial sum accumulation, (c) proposed INT8 additive partial sum quantization ($gs=1$).}
    \vspace{-4mm}  
  \label{figure3}
\end{figure}
where $Q_k(\cdot)$ represents the k-bit quantization function that compresses the high-precision input data to low-bit INT-k with a scaling factor $\alpha$. Specifically, the $Q_k(x)$ constrains $\frac{x}{\alpha}$ within the range $[-2^{k-1}, 2^{k-1}-1]$ or $[0, 2^{k}-1]$ for the signed and unsigned quantization, bounded by $Q_n$ and $Q_p$. The dequantized $\Tilde{x}$ can be retrieved through re-scaling $Q_k(x)$ by $\alpha$, which is determined either using the min-max technique \cite{kim2021bert} or the learnable alternative  \cite{DBLP:conf/iclr/EsserMBAM20}. In this work, we choose the learnable method LSQ \cite{DBLP:conf/iclr/EsserMBAM20} to quantize both weights and activations/PSUMs, among which the scaling factor $\alpha$ of PSUMs are forced to power-of-two format by learning $2^{\lfloor log_2^\alpha\rceil}$ via straight through estimator (STE) \cite{bengio2013estimating, dong2024genetic}. In this way, the multiplication in the re-scaling could be replaced by the hardware-efficient shift operation. 

\begin{figure}[!t]
  \centering
     \captionsetup{skip=1mm}
    \includegraphics[width=0.95\linewidth]{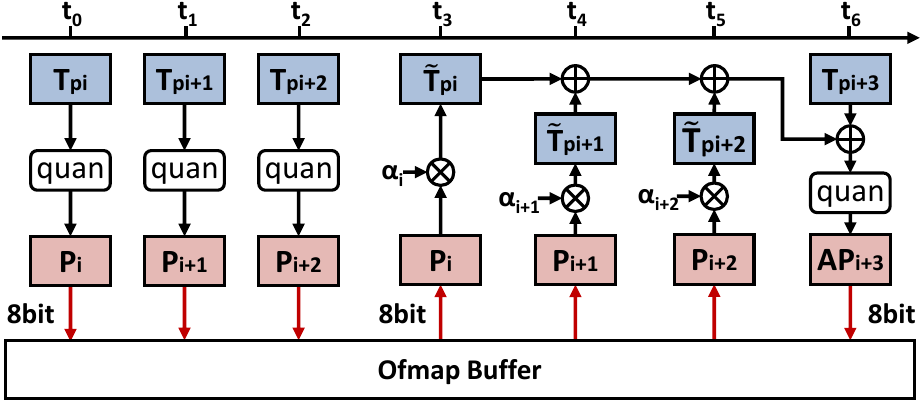}
    \caption{Workflow of additive partial sum quantization with $gs=3$.}
    \vspace{-4mm}  
    \label{Figure4}
\end{figure}

\section{Method}
\subsection{Additive Partial Sum Quantization}
\label{section3.1}
As highlighted in Section \ref{2.1}, tile-based computation (TBC) is a primary method in DNN accelerators due to computational limitations. This technique frequently involves additive interactions with PSUMs to produce the final output tile. Defining $T_i$, $T_w$, $T_p$, and $T_o$ as the tiles for ifmap, weight, PSUM, and ofmap severally, a typical PSUM accumulation example with TBC is illustrated in Fig. \ref{figure3b}. The ofmap tile $T_o$ is derived from the accumulation of multiple PSUMs. Thus, the TBC could be formulated as follows:
\begin{equation}
T_o = \sum_{i=0}^{n_p-1} T_{pi},\enspace n_p = \left\lceil \frac{C_i}{P_{ci}} \right\rceil
\end{equation}
where $n_p$ signifies the number of PSUM tile $T_{pi}$ for $0 \leq i \leq n_p-1$. Although TBC ensures computational efficiency, the necessity for high precision (INT32) in PSUMs can still introduce heavy energy overheads, as the numeric results of PSUM accumulation are generally large in Transformers, which is proved in Section \ref{section4}. Besides, the typical PSUM accumulation could be formulated as follows:
\begin{equation}
\label{eq9}
AP_j = \sum_{i=0}^{j} T_{pi},\enspace T_o =AP_j+\sum_{i=j+1}^{n_p-1} T_{pi}
\end{equation}
where each $AP_j$ represents the additive PSUM after the $j^{th}$ PSUM tile $T_{pj}$. Since the numeric results of PSUM accumulation are generally large in Transformers, they are required to be stored in INT32, which is proved in Section \ref{2.1}. To achieve low-bit memory accesses across all $APs$, we propose an additive PSUM quantization approach termed APSQ, detailed as follows:
\begin{equation}
\label{eq10}
AP_i =Q_k^i(T_{pi}+\alpha_{i-1}\cdot AP_{i-1}),\enspace\text{for}\,1\leq i < n_p.
\end{equation}
In the APSQ, each $AP_i$ is recursively determined by applying $Q_k^i$ to the sum of $T_{pi}$ and the preceding additive PSUM $AP_{i-1}$ which is depicted in Fig. \ref{figure3c}. The initial additive PSUM $AP_{0}$ is set as $Q_k^0(T_{p0})$. The output tile $T_o$ is derived upon completing the quantization of the final PSUM. Consequently, $T_o$ can be succinctly expressed as $AP_{n_p-1}$. In this way, each $APs$ could be accessed by low-bit precision while the quantizer considers both the current high-precision PSUM and its previous $APs$.

\subsection{Grouping Strategy}
\label{section3.2}

\begin{algorithm}[!t]
  \caption{Grouping Strategy}
  \label{algorithm1}
  \begin{algorithmic}[1]
    \Require
      Total number of PSUM tiles $n_p$, PSUM tile set $T_p=[T_{p0},\ldots,T_{pn_p-1}]$, scaling factor set $\alpha=[\alpha_0, \alpha_1, \ldots, \alpha_{n_p-1}]$ and the group size $gs$.
    \Ensure
      Output tile $T_o$.
    \State $\alpha$ is stored in the register list, $n_p>0$ and $AP_{i<0}^*=\bm{0}$
    \State $T_o=\alpha_0 \times Q_k^0(T_{p0})$ \Comment{Initialize $T_o$}
    \For{$i = 0$ to $n_p-1$ with step $gs$}
        \State Read $gs$ previous PSUMs from buffer
        \State $AP_{i-1} = \sum_{i=i-gs}^{i-1} {AP}_{i}^* \times \alpha_i$ \Comment{Accumulation}
        \State ${AP}_{i}^* = Q_k^{i}(AP_{i-1} + T_{pi})$ \Comment{Perform APSQ}
        \State Write ${AP}_{i}^*$ to buffer
    \For{$j = i+1$ to $\min(i + gs - 1, n_p-1)$}
        \If {$j<n_p-1$} 
        \State ${AP}_{j}^* = Q_k^{j}(T_{pj})$ \Comment{Perform PSUM quantization}
        \State Write ${AP}_{j}^*$ to buffer
        \Else \Comment{Final output tile $T_{o}$}
        \State Read $n_p-i+1$ previous PSUMs from buffer
        \small
        \State ${T}_{o}=\alpha_{n_p-1}\times Q_k^{n_p-1}(\sum_{l=i}^{n_p-2} {AP}_{l}^* \times \alpha_l + T_{pn_p-1})$ 
        \normalsize
        \EndIf
    \EndFor
\EndFor
  \end{algorithmic}
\end{algorithm}

However, the accuracy may suffer since each PSUM interacts with multiple quantizers, leading to heavy approximation errors due to repeated rounding during quantization in APSQ. To mitigate this issue, reducing the quantization frequency is essential, as it can minimize the additional rounding error. Building on this insight, we proposed a grouping strategy that combines APSQ with typical quantization detailed in Algorithm \ref{algorithm1}. Specifically, the PSUM tile set is partitioned into groups of size $gs$, where APSQ is applied only once per group, targeting the accumulation of the current input PSUM tile and the previous $gs$ dequantized PSUM tiles. An workflow example of the grouping strategy is depicted in Fig. \ref{Figure4}. Each group of $gs$ quantized PSUMs is stored in the Ofmap Buffer using INT8 precision per PSUM and is sequentially read from the buffer for accumulation, with dequantization performed at INT8 precision per PSUM read. Notably, the grouping strategy maintains the same total memory read and write operations for APSQ with both $gs=1$ and $gs>1$.

\subsection{Reconfigurable APSQ Architecture}
While the grouping strategy offers opportunities to enhance accuracy, an excessively large $gs$ may introduce additional memory overhead. Moreover, different models exhibit varying optimal values for $gs$, as demonstrated in Table \ref{table1}. To support APSQ with various $gs$, we designed a Reconfigurable APSQ Engine (RAE) as depicted in Fig. \ref{Figure2}. The RAE features a PSUM Buffer with four SRAM banks, shifter-based quantization/dequantization modules, adders, and a dedicated RAE controller. RAE's work mode is governed by static encodings $s_0$ and $s_1$, in conjunction with a dynamic encoding $s_2$. Static encodings $s_0$ and $s_1$ are derived from a predefined table based on the group size $gs$, configuring the RAE multiplexers to select appropriate PSUM tiles $P_i$ for different $gs$. The dynamic encoding $s_2$ determines whether an APSQ or a PSUM quantization operation is performed.

For instance, when $gs=1$, $s_0$ is set to 00, $s_1$ is irrelevant, and $s_2$ remains 1. In this mode, each PSUM $P_i$ generated by the PE array prompts the RAE to retrieve the previous INT8 PSUM $P_{i-1}$ from PSUM Bank0, dequantize and accumulate it with $P_i$. The result is quantized via a shift operation and written back to Bank0. When $gs=4$, the configuration adjusts: $s_0$ becomes 10, $s_1$ is set to 1, and $s_2$ dynamically toggles. With $s_2$ set to 0, incoming PSUMs are quantized and stored in the corresponding PSUM Banks without APSQ. After four rounds of PSUM quantization, $s_2$ switches to 1, triggering the simultaneous retrieval of four PSUMs, $P_{i-4}$ to $P_{i-1}$, from all four PSUM Banks. These are processed through a two-stage adder pipeline, accumulated with $P_i$, quantized, and then stored in PSUM Bank3. This flexible architecture allows the RAE to efficiently accommodate various group sizes, effectively replacing conventional PSUM accumulation and storage methods in IS/WS-based DNN accelerators.

\begin{table}[t]
\caption{Accuracy comparison between Baseline and APSQ methods across various models and tasks.}
\centering
\begin{tabular}{c c c p{0.5cm} p{0.5cm} p{0.5cm} p{0.5cm} p{0.5cm}}
\toprule
\textbf{Model} & \textbf{Task} & \textbf{Baseline} & \textbf{gs=1} & \textbf{gs=2} & \textbf{gs=3} & \textbf{gs=4} \\
\midrule
\multirow{6}{*}{BERT-Base} & QNLI & 91.32 & 90.26 & 90.77 & \textbf{91.12} & 91.03 \\
 & MNLI & 84.08 & 82.27 & 83.12 & 83.43 & \textbf{83.54} \\
 & RTE & 74.73 & 74.01 & 74.01 & 73.29 & \textbf{75.81} \\
 & STS-B & 87.89 & 86.94 & 87.31 & 87.60 & \textbf{87.61} \\
 & MRPC & 87.99 & 87.25 & 87.01 & \textbf{87.75} & 87.01 \\
 & CoLA & 53.40 & 50.84 & 51.27 & \textbf{52.59} & 52.36 \\
\midrule
Segformer-B0 & \multirow{2}{*}{ADE20K} & 36.72 & 35.83 & \textbf{36.11} & 35.97 & 35.85 \\
EfficientViT-B1 &  & 39.48 & 37.45 & \textbf{38.65} & 38.41 & 38.47 \\
\bottomrule
\end{tabular}
\label{table1}
\vspace{-2mm}  
\end{table}

\section{Experimental Results}
\label{section4}
\subsection{Experiment Setup}
In this section, we evaluate the model accuracy and hardware performance of the proposed APSQ method across both NLP and CV domains. For the NLP experiments, we adopt the BERT-Base model with an input token length of 128 \cite{devlin2018bert} and benchmark it using the GLUE \cite{wang2018glue} dataset, which includes tasks such as QNLI, MNLI, RTE, STS-B, MRPC, and CoLA. In the CV domain, we focus on semantic segmentation using the ADE20K \cite{zhou2017scene} dataset, comprising over 20,000 images with 150 distinct classes. We experiment with two models at a 512$\times$512 input resolution: Segformer-B0 \cite{xie2021segformer}, which employs a Transformer architecture with vanilla attention, and EfficientViT-B1 \cite{Cai_2023_ICCV}, a lightweight model that integrates convolution with linear attention. All baseline models in our experiments are quantized to W8A8, and APSQ is incorporated into the quantization-aware training (QAT) process, guided by a full-precision teacher model for knowledge distillation. We employ task-specific metrics from the GLUE benchmark for evaluating NLP tasks and the standard mean intersection over union (mIoU) metric to assess segmentation performance. Furthermore, we apply the proposed techniques to a large language model LLaMA2-7B for zero-shot commonsense reasoning evaluation.

The configuration of the analytical DNN accelerator, as shown in Fig. \ref{Figure2}, begins with the allocation of memory buffers: 256KB each for the input feature map (ifmap) and output feature map (ofmap), and 128KB for the weight buffer, while the off-chip DRAM is DDR3. In the MAC array setup, the parallelisms are organized as $P_o=16$, $P_{ci}=8$, and $P_{co}=8$ for BERT-Base, Segformer-B0, and EfficientViT-B1. For large language model (LLM) evaluation, the parallelism settings are adjusted to $P_o=1$, $P_{ci}=32$, and $P_{co}=32$, since the input to the LLM at the decoding stage is a vector, enabling $P_o$ to be set to 1.

\subsection{Accuracy Analysis}

As discussed in Sections~\ref{section3.1} and~\ref{section3.2}, applying APSQ to all PSUMs (equivalent to $gs=1$ grouping) significantly degrades performance due to excessive quantization on each PSUM. To validate this, we evaluated accuracy across different $gs$ values in Table~\ref{table1} under full INT8 quantization. Results show that $gs=1$ causes notable accuracy drops, indicating APSQ's insufficiency alone. Although increasing $gs$ generally helps restore accuracy, the improvements are not strictly monotonic. The performance variation across different models and tasks implies that the intrinsic characteristics of each model affect the effectiveness of APSQ, leading to diverse outcomes for different $gs$ values. These observations underscore the necessity of the RAE, which is designed to adapt to varying $gs$ requirements and optimize accuracy accordingly.

\begin{figure}[!t]
  \centering
     \captionsetup{skip=1mm}
    \includegraphics[width=0.95\linewidth]{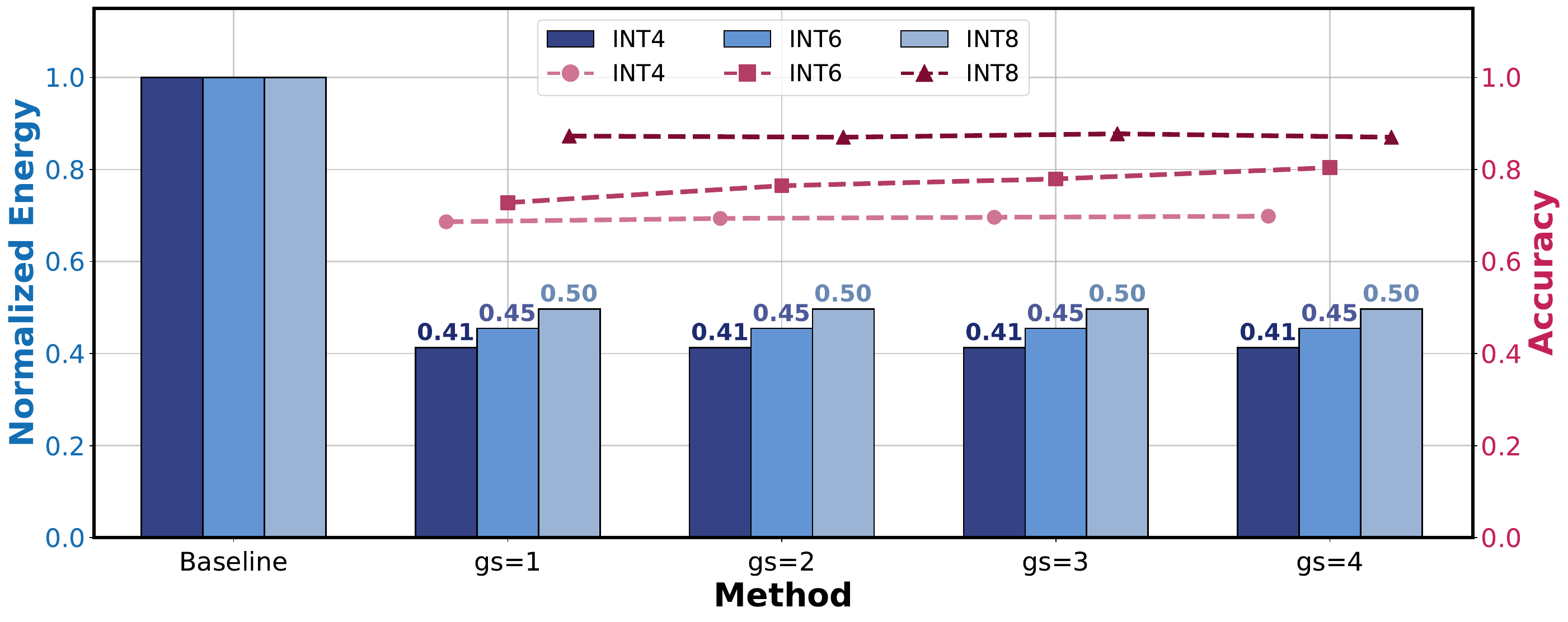}
    \caption{Normalized energy and accuracy across varied $gs$ settings for MRPC under WS dataflow on BERT-Base.}
    \vspace{-4mm}  
    \label{Figure5}
\end{figure}

Additionally, as shown in Fig. \ref{Figure5}, lower PSUM precision consistently reduces energy consumption. However, energy savings weaken below INT8 precision, indicating that reducing PSUM precision further yields minimal benefits and incurs substantial accuracy loss. Hence, adopting INT8 precision for APSQ is technically optimal. Compared to the BERT-Base baseline, the average accuracy metrics determined from the best results of each individual task decrease by only 0.16\%, demonstrating a minimal trade-off. For the semantic segmentation task using Segformer-B0 and EfficientViT-B1, the mIoU decreases by just 0.61\% and 0.83\% with $gs=2$, indicating the INT8 setup with a suitable task-specific $gs$ could strike a balance between accuracy and energy savings.

\subsection{Hardware Performance Analysis}
\vspace{-2mm}  
\begin{table}[h]
\centering
\caption{Hardware synthesis resource consumptions}
\begin{tabular}{l c}
\toprule
 & \textbf{Area ($\mu m^2$)} \\
\midrule
\textbf{Baseline DNN Accelerator} & 1873408 \\
\textbf{RAE} & 86410 \\
\textbf{DNN Accelerator w/ RAE} & 1933674 \\
\bottomrule
\end{tabular}
\label{table2}
\end{table}

\begin{figure}[!t] 
  \centering
    \setlength{\abovecaptionskip}{0.cm}
  \begin{subfigure}{\columnwidth}
    \centering
     \captionsetup{skip=1mm}
    \includegraphics[width=0.95\linewidth]{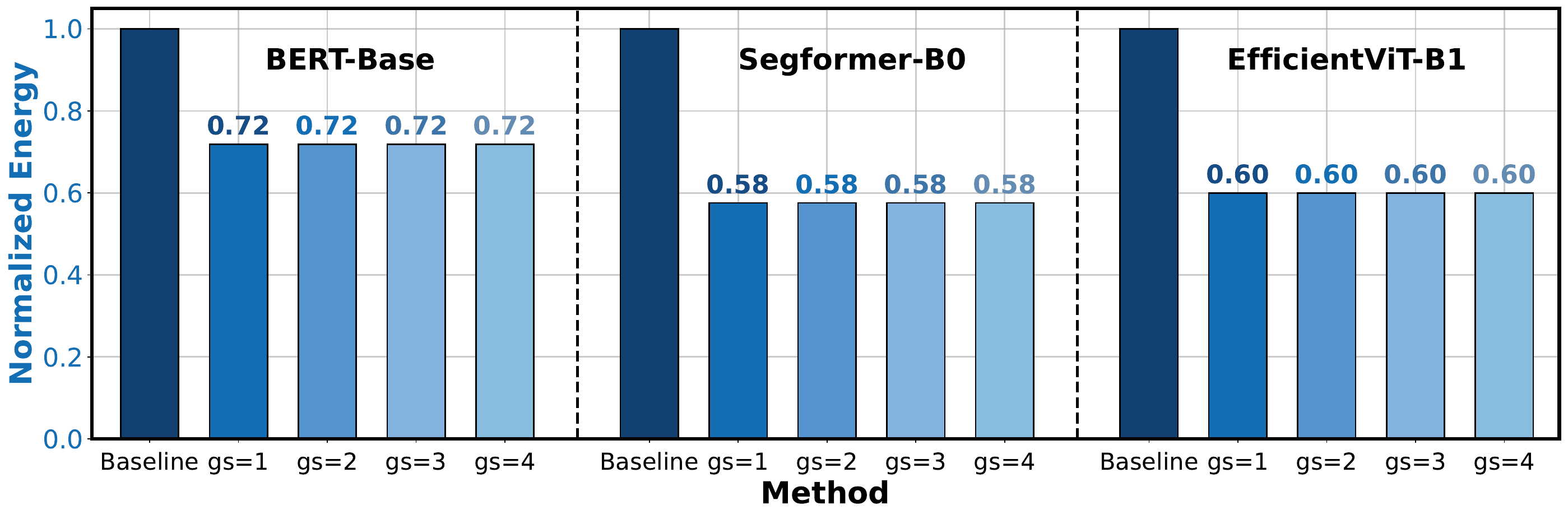}
    \caption{}
    \label{figure6a}
  \end{subfigure}
  \begin{subfigure}{\columnwidth}
    \centering
     \captionsetup{skip=1mm}
    \includegraphics[width=0.95\linewidth]{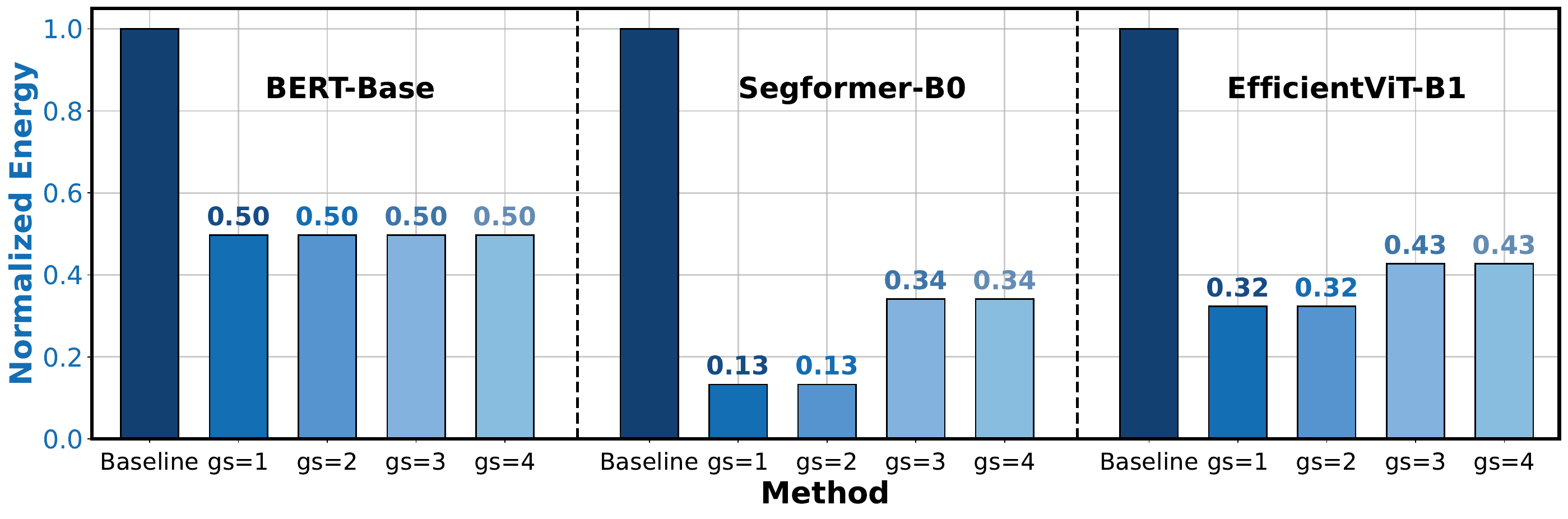}
    \caption{}
    \label{figure6b}
  \end{subfigure}
  \caption{Normalized energy across varied $gs$ settings, and models under (a) IS dataflow, (b) WS dataflow.}
    \vspace{-6mm}  
  \label{figure6}
\end{figure}

In continuation of the prior analyses on accuracy, we evaluate the energy costs associated with different \( gs \) values, following the methods detailed in Section \ref{2.1}. As depicted in Fig. \ref{figure6}, APSQ achieves normalized energy consumption reductions of $28\%$, $42\%$, and $40\%$ for BERT-Base, Segformer-B0, and EfficientViT-B1, respectively, in IS dataflow. The \( gs \) parameter has minimal influence on energy savings for IS, given the typically small PSUM tile size. However, its effect becomes more significant in WS dataflow, where large PSUMs must be buffered. For WS, varying \( gs \) values lead to a uniform $50\%$ energy reduction in BERT-Base, owing to its short token length. In contrast, the impact varies for semantic segmentation models: Segformer-B0 maintains an $87\%$ energy saving at \( gs=1 \) and \( gs=2 \), but this benefit declines to $66\%$ at \( gs=3 \) and \( gs=4 \). Similarly, EfficientViT-B1 achieves a $68\%$ energy reduction for \( gs=1 \) and \( gs=2 \), decreasing to $57\%$ at higher \( gs \) values. This decline occurs as larger PSUM sizes surpass buffer capacity at higher \( gs \), particularly with high-resolution input, resulting in additional DRAM operations.

Consequently, when the PSUM is fully stored on-chip, energy consumption primarily depends on the PSUM precision. However, when the buffer size becomes insufficient, as observed in Segformer-B0 and EfficientViT-B1 under WS, \( gs \) significantly impacts energy consumption. This trade-off between accuracy and energy efficiency emphasizes the necessity of a reconfigurable architecture that can dynamically adjust to various \( gs \) settings, underscoring the essential role of the RAE. We implement a DNN accelerator with and without RAE using Verilog HDL and obtain the hardware resources via Synopsys Design Compiler at 28-nm technology with a frequency constraint of 250 MHz. The accelerator without RAE served as the baseline. As presented in Table \ref{table2}, the accelerator with RAE incurred only a 3.21\% increase in the synthesized area while achieving a 28-87\% reduction in energy consumption as discussed before.

\subsection{Experiment on Large Language Model}
\vspace{-2mm}  

\begin{table}[h]
\caption{Accuracy of Baseline and APSQ across seven Zero-shot Common Sense Reasoning tasks on LLaMA2-7B.}
\centering
\begin{tabular}{c@{\hskip 5pt}c@{\hskip 5pt}c@{\hskip 5pt}c@{\hskip 5pt}c@{\hskip 5pt}c@{\hskip 5pt}c@{\hskip 5pt}c}
\toprule
\textbf{Method} & \textbf{BoolQ} & \textbf{PIQA} & \textbf{HellaS.} & \textbf{WinoG.} & \textbf{Arc-e} & \textbf{Arc-c} & \textbf{OBQA} \\
\midrule
\textbf{Baseline} & 77.80 & 79.22 & 76.64 & 69.69 & 75.25 & 47.10 & 43.40 \\
\textbf{gs=1} & 75.26 & 76.82 & 72.99 & 65.75 & 71.38 & 42.58 & 38.60 \\
\textbf{gs=2} & 75.93 & 77.09 & 74.94 & 67.48 & 73.86 & 46.42 & 42.00 \\
\textbf{gs=3} & 76.45 & \textbf{78.84} & 75.43 & \textbf{68.43} & 73.40 & 47.18 & 41.80 \\
\textbf{gs=4} & \textbf{76.82} & 78.45 & \textbf{76.01} & 67.96 & \textbf{74.75} & \textbf{47.35} & \textbf{42.80} \\
\bottomrule
\end{tabular}
\label{table3}
\vspace{-2mm}  
\end{table}

\begin{table}[h]
\caption{Normalized energy across varied $gs$ settings under IS and WS on LLaMA2-7B.}
\centering
\begin{tabular}{cccccc}
\toprule
\textbf{Dataflow} & \textbf{Baseline} & \textbf{gs=1} & \textbf{gs=2} & \textbf{gs=3} & \textbf{gs=4} \\
\midrule
\textbf{IS} & $1.02\times$ & $1\times$ & $1\times$ & $1\times$ & $1\times$ \\
\textbf{WS} & $31.7\times$ & $1\times$ & $1\times$ & $8.42\times$ & $8.42\times$ \\
\bottomrule
\end{tabular}
\label{table4}
\vspace{-2mm}  
\end{table}
Nowadays, large language models (LLMs) represent a significant advancement in the field of AI. The autoregressive nature of LLMs leads to only one token being inferred at a time during the generation phase. To simulate this scenario while keeping the total number of MAC operations unchanged, we adjust the parallelisms to \( P_o = 1 \), \( P_{ci} = 32 \), and \( P_{co} = 32 \). We implement APSQ within the RoLoRA framework \cite{huang2024rolora} and apply W8A8 quantization for LLaMA2-7B as the baseline. Experiments are conducted on seven Zero-shot Common Sense Reasoning (ZCSR) tasks \cite{gao2021framework}. As shown in Table \ref{table3}, APSQ results in an average accuracy reduction of only 0.59\% when compared to the baseline, evaluated using the best accuracy for each individual task. Despite this small accuracy trade-off, it achieves substantial energy savings of \( 1.02\text{-}31.7\times \) with 4096 sequence length considering both prefilling and decoding stages, as shown in Table \ref{table4}, demonstrating APSQ's significant potential in LLM applications. The minimal enhancement of APSQ on IS is primarily due to the feature map being a vector, which is considerably smaller than weight. Further explorations will be conducted in our future work.

\section{Conclusion}

This work highlights the significant energy overhead from high-precision PSUM accesses in DNN accelerators, often overlooked in prior research. To address this, we first refine an PSUM-precision-aware analytical framework to evaluate the energy consumption of DNN accelerators. Then, we propose a novel APSQ method enhanced by a grouping strategy and a reconfigurable architecture. Experimental results demonstrate the intricate relationship between group size and PSUM quantization overhead. The INT8 APSQ implementation achieves minimal accuracy loss on the GLUE and ADE20K datasets while significantly reducing energy costs. Further experiments also reveal APSQ’s potential for efficient acceleration of LLMs. These findings provide a strategic framework for optimizing DNN accelerators that harness APSQ.

\clearpage

\bibliographystyle{IEEEtran}
\bibliography{IEEEabrv, main}

\end{document}